\documentclass[english,reprint,superscriptaddress,secnumarabic,nofootinbib,nobibnotes]{revtex4-1}
\usepackage[T1]{fontenc}
\usepackage[latin9]{inputenc}
\usepackage{geometry}
\usepackage{xcolor}
\usepackage{amsmath}
\PassOptionsToPackage{normalem}{ulem}
\usepackage{ulem}

\makeatletter

\providecolor{lyxadded}{rgb}{0,0,1}
\providecolor{lyxdeleted}{rgb}{1,0,0}

\DeclareRobustCommand{\lyxsout}[1]{\ifx\\#1\else\sout{#1}\fi}

\setcitestyle{authoryear,round}

\usepackage{amsthm}

\def\frontmatter@abstractheading{}

\renewcommand{\p@subsection}{}
\renewcommand{\p@subsubsection}{}

\usepackage{babel}

\makeatother

\usepackage{babel}
\begin{document}
\title{On Universality of Classical Probability with Contextually Labeled
Random Variables: Response to A. Khrennikov}
\author{Ehtibar N. Dzhafarov}
\thanks{Corresponding author: Ehtibar Dzhafarov, Purdue University, Department
of Psychological Sciences, 703 Third Street West Lafayette, IN 47907,
USA. email: ehtibar@purdue.edu.}
\affiliation{Purdue University}
\author{Maria Kon}
\affiliation{Purdue University}
\begin{abstract}
In his constructive and well-informed commentary, Andrei Khrennikov
acknowledges a privileged status of classical probability theory with
respect to statistical analysis. He also sees advantages offered by
the Contextuality-by-Default theory, notably, that it ``demystifies
quantum mechanics by highlighting the role of contextuality,'' and
that it can detect and measure contextuality in inconsistently connected
systems. He argues, however, that classical probability theory may
have difficulties in describing empirical phenomena if they are described
entirely in terms of observable events. We disagree: contexts in which
random variables are recorded are as observable as the variables'
values. Khrennikov also argues that the Contextuality-by-Default theory
suffers the problem of non-uniqueness of couplings. We disagree that
this is a problem: couplings are all possible ways of imposing counterfactual
joint distributions on random variables that de facto are not jointly
distributed. The uniqueness of modeling experiments by means of quantum
formalisms brought up by Khrennikov is achieved for the price of additional,
substantive assumptions. This is consistent with our view of quantum
theory as a special-purpose generator of classical probabilities.
Khrennikov raises the issue of ``mental signaling,'' by which he
means inconsistent connectedness in behavioral systems. Our position
is that it is as inherent to behavioral systems as their stochasticity.\smallskip{}

\end{abstract}
\maketitle
In our target paper (Dzhafarov \& Kon, 2018) we argued that classical
probability theory (CPT) is a universally applicable mathematical
language, unfalsifiable by empirical phenomena or substantive theories.
We considered several phenomena claimed to pose difficulties for CPT,
and showed that these difficulties only arise if one misidentifies
the random variables used to describe these phenomena. We pointed
out a general way of preventing the possibility of such misidentifications:
contextual labeling of random variables. We contrasted the universality
of CPT with substantive, special-purpose theories aimed at deriving
probability distributions for specific empirical situations. We view
quantum formalisms as such a special-purpose theory (or set of models,
if applied outside quantum mechanics). Khrennikov's (2019) commentary
further strengthens our view of CPT by mentioning that statistical
analysis of data is ``fundamentally based'' on CPT. In the target
paper we mention this too: ``Any prediction derived in a special-purpose
theory {[}...{]} can be (and always is, eventually) fully expressed
in the language of CPT\textcolor{black}{, making it possible to relate
the prediction to empirical data}.'' 

\section{Is CPT inadequate if one sticks to observable events?}

However, Khrennikov still thinks that CPT has problems with certain
empirical situations, such as the double-slit experiment in quantum
physics. He asserts ``the impossibility of defining a Kolmogorov
probability space which is based solely on experimentally verifiable
events.'' Thus, he says, in the double-slit experiment CPT can describe
the situation in each of the contexts defined by which slit is open
and which is closed, but ``it is impossible to represent observables
with respect to these contexts in the same Kolmogorov probability
space without introducing unobservable events.'' Perhaps this comment
is prompted by the fact that in the target paper (and, in greater
detail, in Dzhafarov \& Kujala, 2018) we presented a contextuality
analysis of the double-slit experiment that used unobservable events,
such as ``particle hits the detector having passed through left open
slit.'' This, however, is an optional analysis that in no way replaces
the basic CPT description of this empirical situation. In this basic
description, one considers the random variable 
\begin{equation}
R=\left\{ \begin{array}{l}
+1=\textnormal{the particle hits the detector}\\
-1=\textnormal{the particle did not hit the detector}
\end{array}\right.,\textnormal{ }
\end{equation}
with both values being entirely observable, and their probabilities
being empirically estimable. By the general principle we advocated
in the target paper, this random variable should be labeled by its
context, which is also entirely observable:
\begin{equation}
c=\left\{ \begin{array}{l}
1=\textnormal{both slits are closed}\\
2=\textnormal{only the left slit is open}\\
3=\textnormal{only the right slit is open}\\
4=\textnormal{both slits are open}
\end{array}\right..
\end{equation}
We have therefore four distinct random variables $R^{c}$ ($c=1,\ldots,4$),
stochastically unrelated to each other. Their distributions cannot
be specified unless we make physical assumptions, and it is these
physical assumptions that can agree or disagree with the experiment.
Thus, one can make the physical assumption that 
\begin{equation}
\Pr\left[R^{1}=+1\right]=0,
\end{equation}
i.e., the particle cannot hit the detector if both slits are closed.
It is, as we know, empirically correct, but this is a physical prediction
rather than one of CPT. One can make another physical assumption,
\begin{equation}
\Pr\left[R^{4}=+1\right]=\Pr\left[R^{2}=+1\right]+\Pr\left[R^{3}=+1\right],\label{eq: Feynman}
\end{equation}
and this one is known to be empirically false. This fact was interpreted
by Feynman (1951) as a violation of the additivity law for probabilities,
and Khrennikov seems to concur. We consider this a mistake, because
CPT in no way constrains the probability distributions of $R^{2},R^{3},R^{4}$
in \eqref{eq: Feynman}. The issue, once again, is one of correctly
identifying random variables, and contextual notation (involving,
we emphasize, only observable contexts) is a general way to prevent
mistaken identification.

\section{In CPT, contextual labeling cannot be circumvented}

Khrennikov presents our Contextuality-by-Default (CbD) theory as just
one possible approach within the framework of CPT. We cannot but agree,
if one takes the CbD theory in its entirety. We think, however, that
the departure point of CbD, that all random variables should be labeled
not only by what they measure or respond to but also by the contexts
in which they are recorded, is a logical necessity. 

Khrennikov's exposition of the CbD theory is preceded by an analysis
that contravenes this requirement. He considers a set of random variables
$A_{1},A_{2},B_{1},B_{2}$ such that we have joint distributions for
pairs $\left(A_{i},B_{j}\right)$ ($i,j=1,2$), but not for all four
of them.\footnote{The reader having difficulties in reconciling the notation used in
the present note with that in Khrennikov's commentary (and with the
notation used in the target article) may consult Appendix, where the
correspondences are spelled out. Regrettably, Khrennikov did not keep
his mathematical notation close to one in the target paper, forcing
us here to adopt ``bridging'' notation, both resembling Khrennikov's
and acceptable in the CbD theory, even if not optimal. } He calls these random variables ``observables,'' but it is clear
from Footnote 2 of Khrennikov's commentary that by an ``observable''
Khrennikov means a random variable in the normative CPT sense (as
opposed to, say, Hermitian operators or POVMs in quantum theory, that
are also called ``observables''). However, if $A_{1},A_{2},B_{1},B_{2}$
are classical random variables, the situation considered by Khrennikov
(following many other authors) is logically impossible.

In CPT, random variables are jointly distributed if and only if they
are defined on the same domain probability space. If $X$ and $Y$
are defined on a probability space $\mathcal{S}$, and if $Y$ and
$Z$ are defined on a probability space $\mathcal{S}'$, then $\mathcal{S}'=\mathcal{S}$.
Otherwise, we will have a mathematical impossibility of one and the
same random variable (here, $Y$) defined on two different domains.
But then $X,Y,Z$ are all defined on the same probability space $\mathcal{S}$
and are, therefore, jointly distributed. Applying this simple reasoning
to $A_{1},A_{2},B_{1},B_{2}$, if any three of the pairs
\begin{equation}
\left(A_{1}B_{1}\right),\left(B_{1},A_{2}\right),\left(A_{2},B_{2}\right),\left(B_{2},A_{1}\right)
\end{equation}
are jointly distributed, then all four random variables are jointly
distributed. 

Suppose, however, that Alice records a joint distribution of $\left(X,Y\right)$,
Bob records a joint distribution of $\left(Y,Z\right)$, and Charlie,
who gets the results from both Alice and Bob, finds from some trustworthy
theory that $\left(X,Y,Z\right)$ cannot be jointly distributed. What
should Charlie conclude? The only way of dealing with this situation
is to treat it as a \emph{reductio ad absurdum} case: Charlie has
arrived at a mathematical impossibility, whence there should be a
mistake made in the initial assumptions. If there can be no doubt
that the joint distributions of $\left(X,Y\right)$ and $\left(Y,Z\right)$
exist, and if the ``trustworthy theory'' in question cannot be doubted
either, the only remaining possibility is that it was a mistake to
assume that $Y$ in Alice's $\left(X,Y\right)$ and $Y$ in Bob's
$\left(Y,Z\right)$ is one and the same random variable. Since $\left(X,Y\right)$
and $\left(Y,Z\right)$ are measured separately (in different contexts),
it is perfectly possible that we have in fact $\left(X,Y\right)$
and $\left(Y',Z\right)$ with $Y$ and $Y'$ being distinct random
variables. This would be obvious to Charlie if $Y$ and $Y'$ measured
different things or responded to different questions (in the CbD terminology,
if they had different \emph{contents}). This would also be obvious
to Charlie if the distributions of $Y$ and $Y'$ were different.
However, Charlie can be ``fooled'' into confusing $Y$ with $Y'$
if they have the same content and the same distribution: in this case
the contradiction arrived at by Charlie is the only guide in deciding
that $Y$ and $Y'$ are distinct random variables.

In the CbD theory, the possibility of erroneously assuming $Y=Y'$
is precluded by ``automatically'' defining the identity of a random
variable by both its content and its context. So we have $\left(X_{1}^{1},Y_{2}^{1}\right)$
for Alice, two random variables in the same context $c=1$, distinguished
from each other by their different contents ($q=1$ and $q=2$); and
we have $\left(Y_{2}^{2},Z_{3}^{2}\right)$ for Bob, provided his
$Y$ has the same content ($q=2$) as Alice's $Y$. Now there is no
possibility of a mistake, as $Y_{2}^{1}$ and $Y_{2}^{2}$ are not
the same random variable. Charlie can meaningfully ask, however, whether
$\left(X_{1}^{1},Y_{2}^{1}\right)$ and $\left(Y_{2}^{2},Z_{3}^{2}\right)$
can be coupled in a special way, in particular, if there is a coupling
$\left(\widetilde{X}_{1}^{1},\widetilde{Y}_{2}^{1},\widetilde{Y}_{2}^{2},\widetilde{Z}_{3}^{2}\right)$
in which $\widetilde{Y}_{2}^{1}=\widetilde{Y}_{2}^{2}$ with probability
1. The affirmative answer to this question would mean that the naive
substitution of $\left(X,Y\right)$ and $\left(Y,Z\right)$ for the
rigorously defined $\left(X_{1}^{1},Y_{2}^{1}\right)$ and $\left(Y_{2}^{2},Z_{3}^{2}\right)$
is relatively innocuous: it can be viewed as an informal simplification
that, with some care, leads to no confusion. If, however, a coupling
$\left(\widetilde{X}_{1}^{1},\widetilde{Y}_{2}^{1},\widetilde{Y}_{2}^{2},\widetilde{Z}_{3}^{2}\right)$
with $\widetilde{Y}_{2}^{1}=\widetilde{Y}_{2}^{2}$ does not exist,
then the naive treatment is misleading and confusing.

Returning to the CHSH-related situation focused on by Khrennikov,
in the CbD theory we ``automatically'' have to describe it by eight
random variables grouped into four jointly distributed pairs,
\begin{equation}
\left(A_{1}^{1},B_{1}^{1}\right),\left(B_{1}^{2},A_{2}^{2}\right),\left(A_{2}^{3},B_{2}^{3}\right),\left(B_{2}^{4},A_{1}^{4}\right).\label{eq: Alice-Bob CbD}
\end{equation}
Consider the situation when this system is \emph{consistently connected},
i.e., when the distributions of the content-sharing random variables
(those denoted by the same letter and the same subscript, e.g., $A_{2}^{2}$
and $A_{2}^{3}$) are identical. Rather than ignoring the contexts
(superscripts), which, as we know, may lead to a contradiction, in
the CbD theory we consider this system's couplings

\begin{equation}
\widetilde{A}_{1}^{1},\widetilde{B}_{1}^{1},\widetilde{B}_{1}^{2},\widetilde{A}_{2}^{2},\widetilde{A}_{2}^{3},\widetilde{B}_{2}^{3},\widetilde{B}_{2}^{4},\widetilde{A}_{1}^{4},\label{eq: coupling CbD}
\end{equation}
defined as sets of jointly distributed random variables that respect
all the probabilities that characterize the correspondingly labeled
random variables in \eqref{eq: Alice-Bob CbD}. The question we ask
in the CbD theory is whether among these couplings we can find ones
in which, with probability 1,
\begin{equation}
\widetilde{B}_{1}^{1}=\widetilde{B}_{1}^{2},\widetilde{A}_{2}^{2}=\widetilde{A}_{2}^{3},\widetilde{B}_{2}^{3}=\widetilde{B}_{2}^{4},\widetilde{A}_{1}^{4}=\widetilde{A}_{1}^{1}.\label{eq: equalties CbD}
\end{equation}
Such a coupling is called \emph{maximally connected}, and if it exists,
the system is \emph{noncontextual}. Otherwise it is \emph{contextual}. 

\section{Is non-uniqueness of couplings a problem?}

Khrennikov's account of the CbD theory, using the example \eqref{eq: Alice-Bob CbD}
above, is fairly accurate, except for his assertion that the random
variables in \eqref{eq: Alice-Bob CbD} are jointly distributed. In
the CbD theory, the random variables belonging to different contexts
are stochastically unrelated, i.e., they have no joint distribution.
It is the couplings \eqref{eq: coupling CbD} subject to \eqref{eq: equalties CbD}
that impose all possible joint distributions on the eight random variables
in \eqref{eq: Alice-Bob CbD}. For all practical purposes, however,
Khrennikov's imprecision here is relatively innocuous. Instead of
speaking of four stochastically unrelated pairs of random variables,
Khrennikov speaks of eight jointly distributed variables about which
we only know joint distributions within same-context pairs. The overall
joint distribution then should be treated as unknown (and unknowable,
because different contexts are mutually exclusive). Instead of different
couplings, then, one would speak of different possibilities for this
unknown joint distribution. 

The only problem with this language is that the existence of one ``true''
but unknown joint distribution is disconcerting. What could this distribution
be? Khrennikov indeed finds fault with the CbD theory in this respect.
``How can one select the `right coupling'?'' he asks, and suggests
a substantive interpretation for our inability to find one, in the
spirit of the Copenhagen school of quantum mechanics: perhaps, he
says, it is due to the fact that ``real physical (or psychological)
contextuality is determined not only by semantically defined observables,
but also by apparatuses used for their measurement.'' 

We think that the answer is much simpler: there is no ``right coupling''
because couplings are not part of a system of random variables being
coupled. Couplings are imposed on a system like \eqref{eq: Alice-Bob CbD}
rather than found in it. They formalize the answer to the following
counterfactual question: while in reality the different pairs in \eqref{eq: Alice-Bob CbD}
are stochastically unrelated, what could their joint distribution
be if they were jointly distributed? The set of all possible couplings
of a system characterizes this system, in the same way as the set
of all factors of an integer characterizes this integer. We do not
think of the multitude of the factors as a deficiency of number theory
or our ignorance of one ``true'' factor. One should also keep in
mind that in CbD analysis we are interested not in all possible couplings
but only in special, maximally connected ones. By definition, if a
system is contextual no such couplings exist. So, the non-uniqueness
problem here does not arise. 

Khrennikov contrasts the non-uniqueness of couplings in the CbD theory
with the uniqueness of quantum formalisms: ``{[}Quantum probability{]}
does not suffer the non-uniqueness problem. There is one fixed quantum
state given by a normalized vector $\psi$, or generally by a density
operator $\rho$; and there is the unique representation of observables
by Hermitian operators.'' This is consistent with our view of quantum
formalisms as special-purpose computations aimed at generating probability
distributions, in the classical sense of the latter term. The uniqueness
in question is achieved at the price of making additional assumptions
about a system being studied, leading to empirically falsifiable predictions. 

The following example, a modification of one given in a conference
paper by Robert Griffiths (Griffiths, 2018), may serve as an illustration.
Consider the system
\begin{equation}
\begin{array}{|c|c|c||c}
\hline A_{1} & B_{2} &  & c=1\\
\hline  & B'_{2} & C_{3} & c=2\\
\hline\hline q=1 & q=2 & q=3
\end{array}.\label{eq: E system}
\end{equation}
Irrespective of the row-wise distributions, this system is noncontextual,
and it generally has an infinity of maximally connected couplings.
Suppose now that that $\left\{ A_{1},B_{2},B'_{2},C_{3}\right\} $
are generated by quantum measurements of $q=1,2,3$ in some fixed
quantum state. In particular, invoking the no-disturbance principle,
$B_{2}$ and $B'_{2}$ are identically distributed. Then either the
joint distribution of the four random variables with $B_{2}=B'_{2}$
exists uniquely (if the Hermitian matrices generating $A_{1}$ and
$C_{3}$ commute), or it does not exist (if they do not commute).
We have no multitude of possible joint distributions here, but only
because we have made additional assumptions about the origins of the
random variables.

\section{On ``mental signaling''}

Khrennikov correctly points out that the main advantage of the CbD
conceptualization is its natural extendibility to \emph{inconsistently
connected} systems, i.e., those in which content-sharing random variables
in different contexts may have different distributions. In relation
to the system \eqref{eq: Alice-Bob CbD}, the extended definition
is that the system is noncontextual if it has a coupling \eqref{eq: coupling CbD}
in which each of the equations in \eqref{eq: equalties CbD} holds
with maximal possible probability (constrained by the marginal distributions
of the random variables). This makes the CbD theory especially applicable
to behavioral systems, most if not all of which are inconsistently
connected. Khrennikov uses the term ``mental signaling'' to refer
to this inconsistency, because ``signaling'' is the technical term
used for inconsistent connectedness in some physical systems. He points
out that inconsistently connected systems are also abundant in quantum
physics, for a variety of reasons. (Khrennikov, in fact, was instrumental
in bringing this issue to the attention of experimental physicists.)
However, most physicists would agree that inconsistent connectedness
can be greatly reduced or altogether eliminated by better experimental
designs and greater control over experimental arrangements. In this
relation Khrennikov poses a question of whether or not the same may
be true for the ubiquitous inconsistent connectedness of behavioral
systems. He writes: ``In psychology the situation is more complicated.
There are no theoretical reasons to expect no signaling. Therefore,
it is not obvious whether signaling is a technicality or a fundamental
feature of cognition. {[}...{]} it may be that mental signaling is
really a fundamental feature of cognition.'' 

We think the latter is definitely the case. Psychological experiments
share with quantum physical experiments the fundamental presence of
stochasticity in responses to stimuli (measurements of properties).
Regardless of how strictly these stimuli (properties) are controlled,
stochasticity in the responses (measurements) cannot be reduced by
nontrivial amounts. In addition, however, psychological experiments
exhibit the property of non-selectivity in responses to several stimuli:
when the task is to respond to stimulus $x$ and to respond to stimulus
$y$, then $x$ almost always, if not always, influences the distribution
of responses to $y$, and vice versa (Cervantes \& Dzhafarov, 2018;
Dzhafarov, Zhang, \& Kujala, 2015). In this respect behavioral systems
differ from some quantum physical systems, those in which quantum
theory includes ``no-signaling'' or ``no-disturbance'' constraints.
These constraints, however, do not apply to all quantum physical systems
(see, e.g., a CbD analysis of the Leggett-Garg-type systems in Bacciagaluppi,
2016, and a CbD analysis of the double-slit experiment in Dzhafarov
\& Kujala, 2018).

\section{Conclusion}

Khrennikov's commentary is a constructive and even sympathetic analysis
of our views on CPT with contextual labeling of random variables.
The sympathy in the commentary can be easily understood: Khrennikov
has himself advocated variants of contextual labeling of random variables,
with some publications well preceding the CbD theory (Khrennikov,
2009a, 2009b, 2015a, 2015b). In this note we have discussed those
aspects of Khrennikov's commentary that allowed us to further elucidate
our positions on CPT and the conceptual advantages of the CbD theory.
Khrennikov defends some of the CPT inadequacy claims by bringing up
the observability-of-events restriction. We have disagreed: contexts
are observable too. Khrennikov uses noncontextual indexation of random
variables and introduces contextuality as non-existence of a global
distribution of a set of random variables whose overlapping subsets
are jointly distributed. We have argued that this widespread description
leads to contradictions that can only be avoided by labeling all random
variables contextually. Khrennikov points out a problem of non-uniqueness
faced by the CbD theory in finding couplings for systems of random
variables. We have argued that this is not a problem because the various
joint distributions imposed by couplings do not exist in the same
sense as the random variables being coupled. Rather, the couplings
are counterfactual, describing all possible ways in which variables
that are de facto stochastically unrelated could be related to each
other if they were jointly distributed. Finally, we have supported
Khrennikov's assumption that inconsistent connectedness in behavioral
systems is inherent and irreducible. 

\section*{REFERENCES}

\setlength{\parindent}{0cm}\everypar={\hangindent=15pt}

Bacciagaluppi, G. (2016). Einstein, Bohm and Leggett-Garg. In E. Dzhafarov,
S. Jordan, R. Zhang \& V. Cervantes (Eds.), Contextuality from Quantum
Physics to Psychology (pp. 63-76). New Jersey: World Scientific.

Cervantes, V.H., \& Dzhafarov, E.N. (2018). Snow Queen is evil and
beautiful: Experimental evidence for probabilistic contextuality in
human choices. Decision, 5, 193-204.

Dzhafarov, E.N., \& Kon, M. (2018). On universality of classical probability
with contextually labeled random variables. Journal of Mathematical
Psychology, 85, 17-24

Dzhafarov, E.N., \& Kujala, J.V. (2018). Contextuality analysis of
the double slit experiment (with a glimpse into three slits). Entropy,
20, 278; doi:10.3390/e20040278.

Dzhafarov, E.N., Zhang, R., \& Kujala, J.V. (2015). Is there contextuality
in behavioral and social systems? Philosophical Transactions of the
Royal Society A, 374: 20150099.

Feynman, R.P. (1951). The concept of probability in quantum mechanics.
In J. Neyman (Ed.), Proceedings of the Second Berkeley Symposium on
Mathematical Statistics and Probability (pp. 533-541). Berkeley: University
of California Press.

Griffiths, R.B. (2018, November 9). Quantum measurements and contextuality.
Paper presented at the Purdue Winer Memorial Lectures 2018: Probability
and Contextuality. https:\slash\slash www.purdue.edu\slash hhs\slash psy\slash conferences\slash pwml\slash slides\slash PWML18\_Griffiths.pdf.

Khrennikov, A. (2009a). Contextual Approach to Quantum Formalism.
Dordrecht: Springer.

Khrennikov, A. (2009b). Bell\textquoteright s inequality: Physics
meets probability. Information Science, 179, 492-504.

Khrennikov, A. (2015a). CHSH inequality: Quantum probabilities as
classical conditional probabilities. Foundations of Physics, 45, 711-725.

Khrennikov, A. (2015b). Two-slit experiment: Quantum and classical
probabilities. Physica Scripta, 90, 1-9.

Khrennikov, A. (2019). Classical versus quantum probability: Comments
on the paper \textquotedblleft On universality of classical probability
with contextually labeled random variables\textquotedblright{} by
E. Dzhafarov and M. Kon. 

\bigskip{}
\bigskip{}

\section*{Appendix: Notation}

In the target paper we use the usual CbD notation $R_{q}^{c}$ for
a random variable measuring, or responding to $q=1,2,\ldots$ in context
$c=1,2,\ldots$. Thus, $R_{3}^{2}$, e.g., is a random variable with
content $q=3$ in context $c=2$, and this uniquely identifies it
within a system of random variables being considered. In this note
$R_{q}^{c}$ will be replaced with $A_{q}^{c}$ and $B_{q}^{c}$,
so that the content of a random variable is now identified by both
its subscript and the letter $A$ or $B$. This is done to bring the
notation closer to Khrennikov's use of $A_{qq'}$ and $B_{q'q}$,
when he refers to the CbD theory, to designate random variables recorded
in context that he denotes by $c=\left(a_{q},b_{q'}\right)$. Khrennikov's
own (noncontextual) notation for random variables is $a_{q},b_{q}$
(not to be confused with the use of the same symbols when he refers
to the CbD theory: there $a_{q},b_{q}$ are not random variables).
This may be confusing, but the choice was not ours. In this note,
when we speak of noncontextual notation (considered incorrect in the
CbD theory), we use $A_{q}$, $B_{q}$. 

Khrennikov focuses on systems of random variables that in the CbD
theory are called cyclic systems of rank 4. Their notation choices
here are as shown:

\begin{widetext}

\[
\begin{array}{|c|c|c|c|c}
\hline R_{1}^{1} & {\color{black}R_{2}^{1}} & {\color{black}{\color{blue}}} & {\color{black}{\color{blue}}} & c=1\\
\hline {\color{black}{\color{blue}}} & {\color{black}R_{2}^{2}} & R_{3}^{2} & {\color{black}{\color{blue}}} & {\color{black}c=2}\\
\hline {\color{black}{\color{blue}}} & {\color{black}{\color{blue}}} & {\color{black}R_{3}^{3}} & {\color{black}R_{4}^{3}} & {\color{black}c=3}\\
\hline {\color{black}R_{1}^{4}} & {\color{black}{\color{blue}}} & {\color{black}{\color{blue}}} & {\color{black}R_{4}^{4}} & {\color{black}c=4}\\
\hline {\color{black}q=1} & q=2 & q=3 & q=4 & {\color{black}\begin{array}{c}
\text{target paper,}\\
\textnormal{normative in CbD}
\end{array}}
\end{array}
\]
\[
\begin{array}{|c|c|c|c|c}
\hline A_{11} & {\color{black}B_{11}} & {\color{black}{\color{blue}}} & {\color{black}{\color{blue}}} & C_{11}=\left(a_{1},b_{1}\right)\\
\hline {\color{black}{\color{blue}}} & {\color{black}B_{12}} & A_{21} & {\color{black}{\color{blue}}} & C_{21}=\left(a_{2},b_{1}\right)\\
\hline {\color{black}{\color{blue}}} & {\color{black}{\color{blue}}} & {\color{black}A_{22}} & {\color{black}B_{22}} & C_{22}=\left(a_{2},b_{2}\right)\\
\hline {\color{black}A_{12}} & {\color{black}{\color{blue}}} & {\color{black}{\color{blue}}} & {\color{black}B_{21}} & C_{12}=\left(a_{1},b_{2}\right)\\
\hline {\color{black}a_{1}} & b_{1} & a_{2} & b_{2} & {\color{black}\begin{array}{c}
\text{Khrennikov,}\\
\textnormal{when referring to CbD}
\end{array}}
\end{array}\quad\begin{array}{|c|c|c|c|c}
\hline a_{1} & {\color{black}b_{1}} & {\color{black}{\color{blue}}} & {\color{black}{\color{blue}}} & c=\left(a_{1},b_{1}\right)\\
\hline {\color{black}{\color{blue}}} & {\color{black}b_{1}} & a_{2} & {\color{black}{\color{blue}}} & c=\left(a_{2},b_{1}\right)\\
\hline {\color{black}{\color{blue}}} & {\color{black}{\color{blue}}} & {\color{black}a_{2}} & {\color{black}b_{2}} & c=\left(a_{2},b_{2}\right)\\
\hline {\color{black}a_{1}} & {\color{black}{\color{blue}}} & {\color{black}{\color{blue}}} & {\color{black}b_{2}} & c=\left(a_{1},b_{2}\right)\\
\hline {\color{black}a_{1}} & b_{1} & a_{2} & b_{2} & {\color{black}\begin{array}{c}
\text{Khrennikov's own,}\\
\textnormal{noncontextual}
\end{array}}
\end{array}
\]
\[
\begin{array}{|c|c|c|c|c}
\hline A_{1}^{1} & {\color{black}B_{1}^{1}} & {\color{black}{\color{blue}}} & {\color{black}{\color{blue}}} & c=1\\
\hline {\color{black}{\color{blue}}} & {\color{black}B_{1}^{2}} & A_{2}^{2} & {\color{black}{\color{blue}}} & {\color{black}c=2}\\
\hline {\color{black}{\color{blue}}} & {\color{black}{\color{blue}}} & {\color{black}A_{2}^{3}} & {\color{black}B_{2}^{3}} & {\color{black}c=3}\\
\hline {\color{black}A_{1}^{4}} & {\color{black}{\color{blue}}} & {\color{black}{\color{blue}}} & {\color{black}B_{2}^{4}} & {\color{black}c=4}\\
\hline {\color{black}a=1} & b=1 & a=2 & b=2 & {\color{black}\begin{array}{c}
\text{present note,}\\
\textnormal{acceptable in CbD}
\end{array}}
\end{array}\quad\begin{array}{|c|c|c|c|c}
\hline A_{1} & {\color{black}B_{1}} & {\color{black}{\color{blue}}} & {\color{black}{\color{blue}}} & c=1\\
\hline {\color{black}{\color{blue}}} & {\color{black}B_{1}} & A_{2} & {\color{black}{\color{blue}}} & {\color{black}c=2}\\
\hline {\color{black}{\color{blue}}} & {\color{black}{\color{blue}}} & {\color{black}A_{2}} & {\color{black}B_{2}} & {\color{black}c=3}\\
\hline {\color{black}A_{1}} & {\color{black}{\color{blue}}} & {\color{black}{\color{blue}}} & {\color{black}B_{2}} & {\color{black}c=4}\\
\hline {\color{black}a=1} & b=1 & a=2 & b=2 & {\color{black}\begin{array}{c}
\text{present note,}\\
\textnormal{when referring to Khrennikov's}\\
\textnormal{noncontextual notation}
\end{array}}
\end{array}
\]

\end{widetext}
\end{document}